\documentclass[twocolumn,showpacs,showkeys,preprintnumbers,prd,nofootinbib]{revtex4-1}
\usepackage{comment}
\usepackage{graphics}
\usepackage{graphicx}
\usepackage{bm,natbib}
\usepackage{amsmath}
\usepackage[english]{babel}
\usepackage[T1]{fontenc}
\usepackage{amssymb}
\usepackage{amsfonts}
\usepackage{epsfig}
\usepackage{psfrag}
\usepackage{color}
\usepackage[dvipsnames]{xcolor}
\usepackage{multirow}
\usepackage[colorlinks=true,citecolor=blue]{hyperref}
\usepackage{epstopdf}
\usepackage{mathrsfs}  
\usepackage{etoolbox}

\newcommand{\Lagr}{\mathcal{L}}

\begin{document}
\title{Late-time constraints on modified Gauss-Bonnet cosmology}

\author{Francesco Bajardi}
\email{francesco.bajardi@unina.it}
\author{and Rocco D'Agostino}
\email{rocco.dagostino@unina.it}
\affiliation{Scuola Superiore Meridionale, Largo S. Marcellino 10, 80138 Napoli, Italy.}
\affiliation{Istituto Nazionale di Fisica Nucleare (INFN), Sezione di Napoli, Via Cinthia 9, 80126 Napoli, Italy.}

\begin{abstract}
In this paper, we consider a gravitational action containing a combination of the Ricci scalar, $R$, and the topological Gauss-Bonnet term, $G$. Specifically, we study the cosmological features of a particular class of modified gravity theories selected by symmetry considerations, namely the  $f(R,G)= R^n G^{1-n}$ model. In the context of a spatially flat, homogeneous and isotropic background, we show that the currently observed acceleration of the Universe can be addressed through geometry, hence avoiding \emph{de facto} the shortcomings of the cosmological constant. We thus present a strategy to numerically solve the Friedmann equations in presence of pressureless matter and obtain the redshift behavior of the Hubble expansion rate. Then, to check the viability of the model, we place constraints on the free parameters of the theory by means of a Bayesian Monte Carlo method applied to late-time cosmic observations.  
Our results show that the $f(R,G)$ model is capable of mimicking the low-redshift behavior of the standard $\Lambda$CDM model.
Finally, we investigate the energy conditions and show that, under suitable choices for the values of the cosmographic parameters, they are all violated when considering the mean value of $n$ obtained from our analysis, as occurs in the case of a dark fluid.
 
\end{abstract}


\maketitle

\section{Introduction}
\label{introd}

The milestone for modern cosmology represented by the discovery of the accelerating expansion of the Universe \cite{Riess98,Perlmutter99} has undermined our understanding of the cosmic puzzle over the last two decades. Among the many proposals to explain the observed acceleration, the cosmological constant ($\Lambda$) introduced by Einstein is the simplest attempt able to reproduce the exotic features of the dark energy fluid, which is believed to drive the current cosmic expansion \cite{Peebles03,Copeland06}. However, the resulting scenario, known as the standard $\Lambda$CDM model, is affected by the so-called \emph{fine-tuning problem} resulting from the very large difference between the vacuum energy density predicted from particle physics and its observed value \cite{Weinberg89,Padmanabhan03}. Moreover, a further issue, known as the \emph{coincidence problem}, is due to the fact that the present time turns out to coincide with the only time in the cosmic history when the energy densities of matter and vacuum are of the same order of magnitude \cite{Carroll01}. Therefore, several alternative paradigms have been proposed to address these shortcomings\footnote{In order to heal the cosmological constant problem, a recent study has suggested a mechanism for removing the vacuum energy contribution by means of a phase-transition during the inflationary era \cite{D'Agostino22}.}, such as considering peculiar fluids with negative pressure described in terms of scalar fields \cite{Ratra88,Caldwell98,Zlatev99}, or scenarios aiming to unify different cosmological epochs \cite{Sahni00,Scherrer04,Capozziello06,Anton-Schmidt,D'Agostino22b}.

Nevertheless, the lack of compelling and definitive solutions has naturally led to explore also the possibility that modifications of gravity could be the origin of dark energy. In fact, due to incompatibilities with (and among) observations and issues at the theoretical level, alternatives to Einstein's General Relativity (GR) started being developed, providing possible solutions to yet unsolved issues. In this framework, modifications extending the Hilbert-Einstein action caught much attention, due to their capability of reproducing GR under given limits \cite{Clifton:2011jh, Nojiri:2017ncd}. This is the case of $f(R)$ gravity \cite{Carroll04,Starobinsky07,Nojiri11}, whose gravitational action generalizes the Hilbert-Einstein one by including a generic function of the Ricci scalar curvature, $R$. Thus, as soon as $f(R) = R$, GR is fully recovered. 
The $f(R)$ models, characterized by field equations of the fourth order, can mimic, under suitable forms, the dark energy behavior without resorting to $\Lambda$ \cite{Joyce:2014kja, Koyama:2015vza}. However, no $f(R)$ model is so far capable of fitting all the experimental data at once, or reproducing the whole cosmic history better than the $\Lambda$CDM model \cite{Amendola:2006kh, Sotiriou10}. Moreover, leading to higher-order field equations, some $f(R)$ models exhibit ghosts in their Hamiltonian structure, with the consequence that a self-consistent quantization scheme cannot be pursued \cite{Dolgov:2003px, Faraoni:2006sy}. The main features of $f(R)$ gravity, its applications, and the theoretical structure can be found \emph{e.g.} in \cite{Sotiriou10,DeFelice_review,Capozziello_review} and reference therein. 

Among the extensions of GR, a particular interest has been gained by theories involving the Gauss-Bonnet invariant in the gravitational action \cite{Nojiri:2005jg, Li:2007jm, Elizalde:2010jx, DeFelice:2009aj, Oikonomou:2022ksx, Oikonomou:2021kql, Odintsov:2020vjb}. Specifically, within all the possible combinations of the second-order invariants $R^2$, $R^{\mu \nu} R_{\mu \nu}$ and $R^{\mu \nu \rho \sigma} R_{\mu \nu \rho \sigma}$, with $R_{\mu \nu}$ and $R_{\mu \nu \rho \sigma}$ being the Ricci and the Riemann tensors, respectively, there is a particular linear combination leading to a topological surface in four dimensions. Such a topological surface is the Gauss-Bonnet invariant, defined as $G \equiv R^2- 4 R^{\mu \nu} R_{\mu \nu} + R^{\mu \nu \rho \sigma} R_{\mu \nu \rho \sigma}$. This could be of interest to address issues inherent in GR at different energy scales. More precisely, $G$ naturally emerges in gauge theories of gravity, such as Lovelock \cite{Bajardi:2021hya, Lovelock:1971yv, Mardones:1990qc}, Chern-Simons \cite{Achucarro:1986uwr, Gomez:2011zzd} or Born-Infeld \cite{Leigh:1989jq, Tseytlin:1997csa} gravity. Moreover, its topological nature allows to reduce the order of the equations of motion and simplify the dynamics. However, due to the Gauss-Bonnet theorem, $G$ vanishes identically in three dimensions (or less), while it represents a trivial boundary in four dimensions \cite{Marathe:1989tm}. Therefore, in the latter case, it cannot provide dynamical contributions to the field equations. Nonetheless, a generic function of the Gauss-Bonnet term is trivial in three dimensions (or less), with the consequence that $f(G)$ gravity can be taken into account as a suitable modification of GR in four dimensions, due to its capability of restoring Einstein's theory under particular limits \cite{Bajardi:2020osh}. 

Motivated by the above reasons, in this work we consider a gravitational action constituted by a combination of the Ricci scalar and the Gauss-Bonnet term, leading to the $f(R,G)$ theories. These have been extensively studied in different contexts  \cite{DeFelice,Sadjadi:2010kp,Makarenko:2012gm,DeLaurentis:2013ska,Elizalde:2020zcb,Mustafa:2020jln,deMartino:2020yhq,Nojiri:2022xdo, Akbarieh:2021vhv}, providing interesting results on different scales. In particular, here we study the cosmological dynamics of a subclass of the $f(R,G)$ models, selected by symmetry considerations. Our purpose is to test the viability of such a scenario by means of late-time cosmic observations, and check whether it may represent a suitable alternative to the standard cosmological paradigm. 

The present work is organized as follows. In Sec.~\ref{modified gauss}, we discuss the main properties of Gauss-Bonnet gravity and cosmology, focusing on a particular function selected via the Noether symmetry approach.
In Sec.~\ref{SecObs}, we explore the background cosmological dynamics of the selected model and test its viability through a Bayesian analysis based on Monte Carlo methods applied to late-time cosmic observations, such as Supernovae Ia and observational Hubble data. Moreover, a systematic comparison with the predictions of the standard cosmological paradigm is carried out, along with the analysis of deviations from GR and possible tensions with respect to the most recent findings in the literature.
In Sec.~\ref{sec:energy}, we then study the validity of the energy conditions and the physical implications resulting from possible violations of them in terms of the free parameters of the model.
Finally, in Sec.~\ref{conclSec}, we discuss our results, remarking on the main theoretical features exhibited by the model. We thus conclude this work by outlining the future perspectives of the modified Gauss-Bonnet dark energy scenario.

\section{Modified Gauss-Bonnet gravity and cosmology}
\label{modified gauss}
One of the most general extensions of the Hilbert-Einstein action can be built by means of higher-order curvature invariants and dynamical scalar fields, $\phi$, non-minimally coupled to geometry. For instance, one could consider the action\footnote{In this paper, we consider units where $8\pi  G_N=c=\hbar=1$.} 
\begin{eqnarray}
S = \int  d^4x\, \sqrt{-g}\, && f \Big(\phi, R, \Box R, ..., \Box^{n} R, \nonumber 
\\
&& \hspace{0.6cm} R^{\mu \nu}R_{\mu \nu}, R^{\mu \nu \rho \sigma} R_{\mu \nu \rho \sigma}\Big) \,,
\label{HOaction}
\end{eqnarray}
containing higher-order derivatives in the geometric terms and leading to $2n+4$-th order field equations. Here, $g$ is the determinant of the metric tensor $g_{\mu\nu}$, whereas $\Box \equiv \nabla_\mu \nabla^\mu$ is the D'Alembert operator, with $\nabla_\mu$ being the covariant derivative. 

As previously mentioned, we shall focus on a particular subcase of the action \eqref{HOaction}, containing a function of the scalar curvature and the Gauss-Bonnet invariant. In particular, by defining $P \equiv R^{\mu \nu} R_{\mu \nu}$ and $\mathcal{Q} \equiv R^{\mu \nu \rho \sigma} R_{\mu \nu \rho \sigma}$, the variation of the action
\begin{equation}
S = \int  d^4 x\, \sqrt{-g} \, f(R,P,\mathcal{Q}) \,,
\label{fpqact}
\end{equation}
yields the following field equations \cite{Carroll:2004de, Bogdanos:2009tn}:
\begin{align}
& f_R \left(R_{\mu\nu}-\frac{1}{2}g_{\mu\nu}R\right) = \frac{1}{2} g_{\mu \nu} f - \left(R+g_{\mu \nu} \Box - \nabla_\mu \nabla_\nu \right)f_R  \nonumber 
\\
&- 2 \left(f_P R^\alpha_\mu R_{\alpha \nu} + f_{\mathcal{Q}} R_{\rho \sigma \alpha \mu} R^{\rho \sigma \alpha}_{\;\;\;\;\;\; \nu} \right)-g_{\mu \nu} \nabla_\rho \nabla_\sigma \left(f_P R^{\rho \sigma}\right)\nonumber
\\
&-  \Box \left(f_P R_{\mu \nu} \right)  
+ 2 \nabla_\sigma \nabla_\rho \left[f_P R^\rho_{\left\{ \mu \right.} \delta^\sigma_{ \nu \left. \right\}} + 2 f_{\mathcal{Q}} R^{\rho\;\;\;\;\; \sigma}_{\; \left\{ \mu \nu\right\}} \right],  
\label{field equations f(R,P,Q)}\, 
\end{align}
where $ \{ \} $ denotes the anti-commutator, while we have defined
\begin{equation}
f_R\equiv \frac{\partial f}{\partial R}\,, \quad f_P\equiv \frac{\partial f}{\partial P}\,, \quad f_\mathcal{Q}\equiv \frac{\partial f}{\partial \mathcal{Q}}\,.
\end{equation}
The Gauss-Bonnet topological invariant arises when considering the combination $f(R, P, \mathcal{Q}) = f(R, R^2 -4P+ \mathcal{Q}) \equiv f(R, G)$. Under this assumption, one obtains the action
\begin{equation}
S= \int d^4x\, \sqrt{-g}\, \left[f(R, G)+\mathcal{L}_m\right]  ,
\label{actionfRG}
\end{equation}
and the field equations become
\begin{align}
&f_R \left(R_{\mu\nu}-\frac{1}{2}g_{\mu\nu}R\right) = \frac{1}{2} g_{\mu \nu}\left(f-R f_{R}\right)+ \nabla_{\mu} \nabla_{\nu} f_{R} 
\\
&-g_{\mu \nu} \square f_{R}  + f_{G} \left(-2 R R_{\mu \nu}+4 R_{\mu \rho} R_{\nu}^{\rho}-2 R_{\mu}^{\,\, \rho \lambda \sigma} R_{\nu \rho \lambda \sigma} \right.  \nonumber
\\
&\left. +4 g^{\rho \lambda} g^{\sigma \alpha} R_{\mu \rho \nu \sigma} R_{\lambda \alpha}\right) +2\left(\nabla_{\mu} \nabla_{\nu} f_{G}\right) R - 2 g_{\mu \nu} (\square f_{G}) R \nonumber
\\
&+ 4\left(\square f_{G}\right) R_{\mu \nu}-4\left(\nabla_{\rho}\nabla_\mu f_{G}\right) R_{\nu}^{\rho}  -4\left(\nabla_\rho \nabla_\nu f_{G}\right) R_{\mu}^{\rho} \nonumber
\\
&  +4 g_{\mu \nu}\left(\nabla_\rho \nabla_\lambda f_{G}\right) R^{\rho \lambda} -4\left(\nabla_{\lambda} \nabla_{\alpha} f_{G}\right) g^{\rho \lambda} g^{\sigma \alpha} R_{\mu \rho \nu \sigma} + T_{\mu \nu} , \label{f(RG)FE} \nonumber
\end{align}
where $f_{G}\equiv \frac{\partial f}{\partial G}$ and $T_{\mu \nu}$ is the energy-momentum tensor associated to the matter Lagrangian density $\mathcal{L}_m$, namely
\begin{equation}
T_{\mu\nu}= -\dfrac{2}{\sqrt{-g}}\dfrac{\delta \mathcal{L}_m}{\delta g^{\mu\nu}}\,.
\end{equation}

Interestingly, within the cosmological context, it turns out that the function $f(G) \sim \sqrt{G}$ behaves like the scalar curvature, thus permitting to recover the Einstein-Hilbert action even without imposing the GR limit as a requirement \cite{Bajardi:2020osh}.
Therefore, the introduction of $G$ can play the role of an effective cosmological constant given by curvature. Nonetheless, as pointed out in \cite{DeFelice:2009ak}, higher-order derivatives can induce the presence of superluminal ghosts at the level of cosmological perturbations. This causes the impossibility of recasting the Lagrangian into a canonical form, so that the Hamiltonian becomes linearly unstable. However, in~\cite{Astashenok:2015haa, Nojiri:2018ouv}, the authors show that the Lagrange multipliers can, in principle, address this issue leading to ghost-free primordial curvature perturbations. This can be proved by casting $f(R,G)$ gravity in the Jordan frame, thus coupling the Gauss-Bonnet invariant with a dynamical scalar field and choosing a suitable form for the resulting extra potential.

To explore the cosmological dynamics of $f(R,G)$ gravity, let us consider the spatially-flat Friedmann-Lema\^itre-Robertson-Walker (FLRW) line element
\begin{equation} 
ds^2=-dt^2+a(t)^2\delta_{ij}dx^idx^j\,,
\label{linelement}
\end{equation}
where $a(t)$ is the scale factor\footnote{We here follow the standard recipe, according to which the scale is normalized to the unity at the present time.} depending on cosmic time, $t$.
Hence, the Gauss-Bonnet scalar can be expressed as
\begin{equation}
G  = 24 \left(\frac{\dot{a}^2 \ddot{a}}{a^3} \right)= \frac{8}{a^3}\frac{d}{dt} \left(\dot{a}^3 \right),
\label{GBEXPR}
\end{equation}
from which one can notice that the quantity $\sqrt{-g} \, G$ is a total derivative. 
Moreover, neglecting radiation and assuming pressureless matter, the modified Friedmann equations  read
\begin{align}
&H^2=\frac{1}{3}\left(\dfrac{\rho_m}{f_R}+\rho_{de}\right), \label{eq:first} \\
&2\dot{H}+3H^2=-p_{de}\,, \label{eq:second}
\end{align}
where
\begin{align}
\rho_{de}&=\frac{1}{2f_R}(Rf_R+G f_G-f-6H \dot{f_R} -24 H^3\dot{f_G}) \label{rde}
\,,\\
p_{de}&=\dfrac{1}{f_R}\Big[2H\dot{f_R}+\ddot{f_{R}}+8H^3\dot{f_G}+8H\dot{H} \dot{f_G}+4H^2\ddot{f_G} \nonumber \\
&\hspace{1.2cm} -\frac{1}{2}(Rf_R+Gf_G-f)\Big],
\label{pde}
\end{align}
and
\begin{align}
R&=6(2H^2+\dot H)\,, \label{cosmoexprR}\\ 
G&=24H^2(H^2+\dot H)\,\label{cosmoexprG}.
\end{align}

In the Lagrangian formalism, it is possible to use the cosmological expressions of $R$ and $G$ as Lagrange multipliers and obtain the point-like Lagrangian. Specifically, when considering the line element \eqref{linelement}, the action \eqref{actionfRG} can be written as
\begin{equation}
    S = \int dt \left[a^3 f - \lambda \left(R - 6 \frac{\ddot{a}}{a} - 6 \frac{\dot{a}^2}{a^2} \right) - \tau \left(G - 24 \frac{\ddot{a} \dot{a}^2}{a^2} \right) \right],
\end{equation}
where $\lambda$ and $\tau$ are the Lagrange multipliers. As shown in \cite{Acunzo:2021gqc, Bajardi:2021tul}, the variational principle with respect to $R$ and $G$ can provide the value of $\lambda$ and $\tau$, respectively. Therefore, after integrating out second derivatives, the Lagrangian takes the form
\begin{align}
\mathcal{L}=&\  6 a \dot{a}^{2} f_R +6 a^{2} \dot{a} \left(f_{RR} \dot{R} + f_{R G} \dot{G} \right) \label{lagr f(R,G)} \\
&- 8 \dot{a}^{3} \left(f_{R G} \dot{R} + f_{G G} \dot{G} \right)+a^{3} \left(f -R f_R - G f_G \right). \nonumber
\end{align}
Notice that Eqs.~\eqref{cosmoexprR} and \eqref{cosmoexprG} can be also obtained by the energy condition and the Euler-Lagrange equation with respect to the scale factor, respectively. The former is a condition of zero energy that allows recovering the modified first Friedmann equation when the Lapse function is not included in the starting line element. Moreover, the Euler-Lagrange equations with respect to $R$ and $G$ give back the cosmological expressions of the two scalars by construction. 

\subsection{Selecting $f(R,G)$ models by Noether symmetries}

Sketching the steps reported in~\cite{Capozziello:2014ioa, Camci:2018apx}, we here show how to select viable $f(R,G)$ models by means of the so-called \emph{Noether symmetry approach} (see \cite{Bajardi:2020xfj, Urban:2020lfk, Dialektopoulos:2018qoe, Bajardi:2022ypn} for details). To do this, let us first recall that, if $X$ is the generator of a certain transformation being a symmetry for the Lagrangian $\Lagr$ and $X^{[1]}$ its first prolongation, then the following condition must hold:
\begin{equation}
    X^{[1]} \Lagr + \dot{\xi} \Lagr = \dot{\mathfrak{g}}\,,
    \label{Noethcond}
\end{equation}
where $\mathfrak{g}$ is a gauge function depending on the minisuperspace variables. In a generic minisuperspace of the form $\mathcal{S} = \{q^i\}$, the first prolongation of $X$ reads
\begin{equation}
    X^{[1]} = \xi \frac{\partial}{\partial t} + \eta^i \frac{\partial}{\partial q^i} + (\dot{\eta}^i - \dot{\xi} \dot{q}^i) \frac{\partial}{\partial \dot{q}^i}\,,
\end{equation}
with $\eta^i$ and $\xi$ being the infinitesimal generators related to variable transformations and time translations, respectively. Generally, $t$ accounts for an affine parameter, which in cosmology is represented by the cosmic time. In our case, the minisuperspace is made of three variables, namely $\mathcal{S} \equiv \{a,R,G \}$, and the infinitesimal generator $\eta^i$ can be thus decomposed as $\eta^i = \{\alpha, \beta, \zeta\}$. Under these conditions, the Noether vector $X^{[1]}$ becomes
\begin{eqnarray}
X^{[1]} &=& \xi(a,R,G, t) \partial_t + \alpha(a,R,G, t) \partial_a + \beta(a,R,G, t) \partial_R \nonumber
\\
&+& \zeta(a,R,G, t) \partial_{G} + \dot{\alpha}(a,R,G, t) \partial_{\dot{a}} + \dot{\beta}(a,R,G, t)  \partial_{\dot{R}}  \nonumber
\\
&+& \dot{\zeta}(a,R,G, t) \partial_{\dot{G}}\,,
\end{eqnarray} 
and the identity \eqref{Noethcond} applied to the Lagrangian \eqref{lagr f(R,G)} provides a system of 10 differential equations. The selected functions are
\begin{subequations}
\begin{eqnarray}
&& f(R,G) = f_0 R + f_1 G^\frac{3}{2}\,,
\\
&& f(R,G) = f_0 R^\frac{7}{8} + f_1 G\,,
\\
&& f(R,G) = f_0 R^{\frac{1}{2}} + f_1 G^{\frac{1}{4}}\,,
\\
&& f(R,G) = f_0 R^n G^{m}\,.
\end{eqnarray}
\end{subequations}
In what follows, we focus our attention on the latter function and investigate its cosmological properties.
\subsection{The case $f(R,G)= R^n G^{m}$}

Let us then consider the model $f(R,G)=R^n G^m$. To determine the cosmological dynamics, we can make use of the relations reported in Appendix~\ref{app:relations}. Also, one may introduce the cosmographic parameters $(q,j,s)$ \cite{Weinberg72,Visser05,rocco_chebyshev}, and express the time derivatives of the Hubble parameter as follows:
\begin{subequations}
\begin{align}
\dot H&=-H^2(1+q) \,,\\
\ddot{H}&=H^3(j+3q+2)\,,\\
\dddot{H}&=H^4\left[s-4 j-3 q (q+4)-6)\right]\,.
\end{align}
\end{subequations}
Thus, we find
\begin{equation}
\rho_{de}=\dfrac{3H^2}{n(q-1)q^2}\sum_{k=0}^4b_k q^k\,,
\label{eq:rho_de}
\end{equation}
where we have defined $b_k\equiv b_k(j;n,m)$ as
\begin{subequations}
\begin{align}
b_0=&\ m (m -1)  j\,, \\
b_1=&\ m\left[2 n +3 m -2 j (n +m -1)-3\right], \\
b_2=&\ n \left[2 m  (j-2)-j+1\right]+ n ^2 (j-2)\nonumber \\
&+(m -1) \left[m  (j-4)-1\right], \\
b_3=&\ n(3-n)-(m -2) (m -1)\,, \\
b_4=&\ (2 m -1) (n +m -1)\,.
\end{align}
\end{subequations}
It is worth to note that, for $n=1$ and $m=0$, \emph{i.e.} $f(R,G)\rightarrow R$, Eq.~\eqref{eq:rho_de} identically vanishes and one recovers the behavior of pure GR.
Moreover, we find
\begin{equation}
p_{de}=\dfrac{H^2}{n(q-1)^2 q^3}\sum_{k=0}^7c_k q^k\,,
\label{eq:p_de}
\end{equation}
where the lengthy expressions of $c_k\equiv c_k(j,s;n,m)$ are reported in Appendix~\ref{app:p_de}. 

To simplify the calculations and reduce the number of degrees of freedom, we shall consider the case $m=1-n$, namely $f(R,G)= R^n G^{1-n}$, for which the vacuum field equations admit exact solutions \cite{Capozziello:2014ioa}. Clearly, for $n=1$ GR is fully recovered, while for $n=0$ the model only leads to trivial dynamics. The cosmological features of this model have been addressed in several contexts. For instance, the dynamical analysis pursued in~\cite{SantosDaCosta:2018bbw} showed that two (out of eight) fixed points yield possible candidates for the dark energy era, thus predicting the accelerating behavior of the late-time Universe. The authors also showed that for $n > 1$ the Universe undergoes a never-ending acceleration without the possibility of structure formation.
Moreover, in~\cite{Bamba:2010wfw}, a possible way to cure the finite-time future singularities was addressed by higher-order curvature corrections arising from higher-order field equations.
Also, the power-law inflation and the primordial power spectrum were analyzed in \cite{DeLaurentis:2015fea}.
In~\cite{Bahamonde:2019swy}, the Noether symmetry approach was applied to $f(R,G)$ gravity and exact solutions are provided in a static and spherically symmetric background.

Therefore, for $m=1-n$, it is straightforward to show that the effective dark energy density and pressure are given by, respectively,
\begin{align}
\rho_{de}&=\frac{3 H^2 (n-1) \left(j-2 q^3-2 q^2+q\right)}{(q-1) q^2}\,,
\label{rhode} \\
p_{de}&=\frac{H^2 (n-1)}{(q-1)^2 q^3}\Big\{q \big[(4 n-6) q^5+2 (4 n-7) q^4-4 n q^2\nonumber \nonumber \\
& \hspace{2.8cm} +q (n-s-1)+3 q^3+s\big]   \nonumber \\
& \hspace{2.4cm}+2 j q \left(-2 n q^2-2 n q+n+6 q^2\right)\nonumber \\
& \hspace{2.4cm}+ j^2 (n-3 q+1)\Big\}\,.
\label{pide}
\end{align}
Hence, the equation of state (EoS) parameter for dark energy, $w_{de}\equiv p_{de}/\rho_{de}$, reads
\begin{widetext}
\begin{equation}
w_{de}=\dfrac{3 q j^2-j^2-12 j q^3-n  \left(j-2 q^3-2 q^2+q\right)^2+6 q^6+14 q^5-3 q^4+q^2 s+q^2-q s}{3 q \left(j+q-jq-3q^2+2q^4\right)}\,.
\end{equation}
\end{widetext}

Interesting properties can be obtained by considering the field equations in vacuum. Indeed, setting $\rho_m=0$ in Eqs.~\eqref{eq:first} and \eqref{eq:second}, one can obtain analytical solutions (see Appendix~\ref{app:vacuum solutions}). These can be easily handled by introducing the variable $\gamma = R/G$. 
Hence, for $m=1-n$, the point-like Lagrangian can be written as
\begin{equation}
 \Lagr =  2f_0 n \gamma^{n-2} \dot{a} \left[-3 a \dot{a} \gamma - 3(n-1) a^2 \dot{\gamma} + 4 (n-1) \dot{a}^2 \gamma \dot{\gamma}\right].  
\end{equation}
In this way, $G$ turns out to be cyclic and the field equations admit the exact solution
\begin{equation}
    a(t) = a_0 t^{2n-1}\,, \qquad \gamma = \frac{4n-3}{8(n-1)(2n-1)^2} \, t^2\,.
    \label{ExactSol}
\end{equation}

\section{Comparison with observations}
\label{SecObs}
In this section, we shall test the observational viability of the model under study by means of a Bayesian analysis of the late-time cosmic data. In particular, we consider the measurements from the Supernovae (SN) Ia Pantheon catalog \cite{Scolnic18} and the cosmic chronometers (CC) given by the observational Hubble data collected in \cite{Capozziello18}. In fact, statistical analyses based on these datasets allow obtaining reliable outcomes that are not affected by assumptions of any underlying fiducial model \cite{D'Agostino18,D'Agostino19}. In what follows, we describe the main features of such measurements, together with the corresponding Likelihood functions.

\subsection{Supernovae Ia}

The Pantheon sample \cite{Scolnic18} consists of 1048 measurements of SN Ia in the redshift\footnote{The redshift $z$ is related to the scale factor through $z= a^{-1}-1$.}  range [0.01,\,2.3]. In such a catalog, the standardization of each SN is obtained by adopting the SALT2 light-curve fitter\footnote{We refer the reader to \cite{Betoule14} for the details on the parametrization of the SN distance modulus in terms of the light-curve coefficients and the host-galaxy corrections.} \cite{Guy07}.

In the present study, we use the 6 measurements of the quantity $E^{-1}(z)\equiv H(z)/H_0$ as presented in \cite{Riess18}, where $H_0$ is the Hubble constant. These constitute a self-consistent and model-independent set built upon the full Pantheon collection, relying only on the assumption of a spatially flat universe. The Likelihood of the SN data can be thus written as
\begin{equation}
\mathscr{L}_\text{SN}\propto \exp\left\{-\frac{1}{2}\mathbf{v}^\text{T} \mathbf{C}_\text{SN}^{-1} \mathbf{v}\right\} ,
\end{equation}
where deviations from the theoretical expectations are accounted for through the differences $v_i= E^{-1}_{obs,i}-E^{-1}_{th,i}$ evaluated at each data point, while $^\text{T}$ indicates the transpose of the same vector. Moreover, $\mathbf{C}_\text{SN}^{-1}$ is the inverse of the covariance matrix measuring the correlations among the SN data, as reported in \cite{Riess18,DAgostino:2022tdk}.

\subsection{Cosmic Chronometers}

The additional dataset we utilize in our analysis is based on the differential age method \cite{Jimenez02}. The latter permits to investigate the cosmic expansion in a model-independent way through the spectroscopic age measurement of couples of passively-evolving galaxies, which can be thought as chronometers for measuring the redshift variation with respect to the cosmic time, $dz/dt$. Thus, one can obtain the value of the Hubble parameter  from the relation $H(z)=-(1+z)^{-1}dz/dt$.

Specifically, in our study, we take into account the 31 data points up to $z\sim 2$ previously collected in \cite{Capozziello18}\footnote{See also references therein.}. As these measurements are uncorrelated among themselves, we can write the corresponding Likelihood simply as 
\begin{equation}
\mathscr{L}_\text{CC}\propto\exp\left\{-\dfrac{1}{2}\sum_{i=1}^{31}\left(\dfrac{H_{obs,i}-H_{th,i}}{\sigma_{H,i}}\right)^2\right\}  ,
\end{equation}
with $\sigma_H$ being the relative $65\%$ uncertainties associated to the observed $H$ values, $H_{obs}$.

\subsection{Monte Carlo analysis}

The low-redshift data described above can be thus used to place observational constraints over the free parameters of the $f(R,G)=R^n G^{1-n}$ model. To this aim, we adopted the Markov Chain Monte Carlo (MCMC) method by means of the Metropolis-Hasting algorithm \cite{Hastings70} applied to the joint Likelihood, given by
\begin{equation}
\mathscr{L}_\text{joint}=\mathscr{L}_\text{SN}\times \mathscr{L}_\text{CC}\,. 
\end{equation}

The theoretical values of the Hubble rate can be obtained by numerically solving Eq.~\eqref{eq:first}. Assuming matter to behave as a pressureless perfect fluid, we can write $\rho_m=3H_0^2\Omega_{m0}(1+z)^3$, with $\Omega_{m0}$ being the current value of the matter density parameter\footnote{The subscript ``0" refers to quantities evaluated at $z=0$, corresponding to the present time.}.
Thus, for the specific model under consideration, the first Friedmann equation takes the form
\begin{align}
H^2=&\ \dfrac{H_0^2 4^{n-1}\Omega_{m0} (z+1)^3}{n}\left\{\frac{H^2 \left[H-(z+1) H'\right]}{2 H-(z+1) H'}\right\}^{n-1}\nonumber \\
&+\frac{H^2 (n-1) (z+1)}{\left[H-(z+1) H'\right]^2 \left[2 H-(z+1) H'\right]}\left\{2 (z+1)^2 {H'}^3 \right. \nonumber \\
&\left. + H^2 \left[3 H'-(z+1) H''\right] -5 H (z+1) {H'}^2 \right\},
\label{eq:diff}
\end{align}
where the prime denotes the derivative with respect to $z$.
The above equation has been obtained by converting the time derivatives into derivatives with respect to the redshift according to
\begin{equation}
\frac{d}{dt}= -(1+z)H(z) \dfrac{d}{dz}\,.
\end{equation}
Eq.~\eqref{eq:diff} represents a second-order differential equation for the function $H(z)$, which can be solved by means of suitable boundary conditions. The first initial condition is simply $H(0)=H_0$. To determine the second initial condition, one may require that, at the present time, the first derivative of the Hubble parameter agrees with the predictions of the standard $\Lambda$CDM model, which is characterized by the following expansion law:
\begin{equation}
H_{\Lambda\text{CDM}}=H_0 \sqrt{\Omega_{m0}(1+z)^3+1-\Omega_{m0}}\,.   
\end{equation}
Thus, taking the first derivative of the above equation with respect to $z$, one finds
\begin{equation}
H_{\Lambda\text{CDM}}'=\frac{3 H_0 \Omega_{m0} (1+z)^2}{2 \sqrt{\Omega_{m0} (1+z)^3+1-\Omega_{m0}}}\,,
\end{equation}
which determines the second initial condition for Eq.~\eqref{eq:diff}, namely $H'(0)=3H_0\Omega_{m0}/2$. 

In our numerical analysis, we considered the reduced Hubble constant $h\equiv H_0/(100$ km $s^{-1}$ Mpc$^{-1}$), which represents a free parameter of the model, together with $\Omega_{m0}$ and $n$. We thus assumed the cosmological parameters as uniformly distributed within the following ranges:
\begin{equation}
h\in (0.5,0.9)\,, \quad \Omega_{m0}\in (0,1)\,, \quad n\in (1,2)\,.
\end{equation}
In order to constrain the cosmological parameters, we ran a small initial chain of 2,000 steps, from which we removed the first 100 ones to account for the burn-in phase. This provided us with a test covariance matrix that served as a starting guess for the subsequent main chains. We then ran five independent chains of 20,000 steps each, which have been eventually merged into a final bigger chain of 1,000,000 points.

\begin{figure}
    \centering
    \includegraphics[width=3.4in]{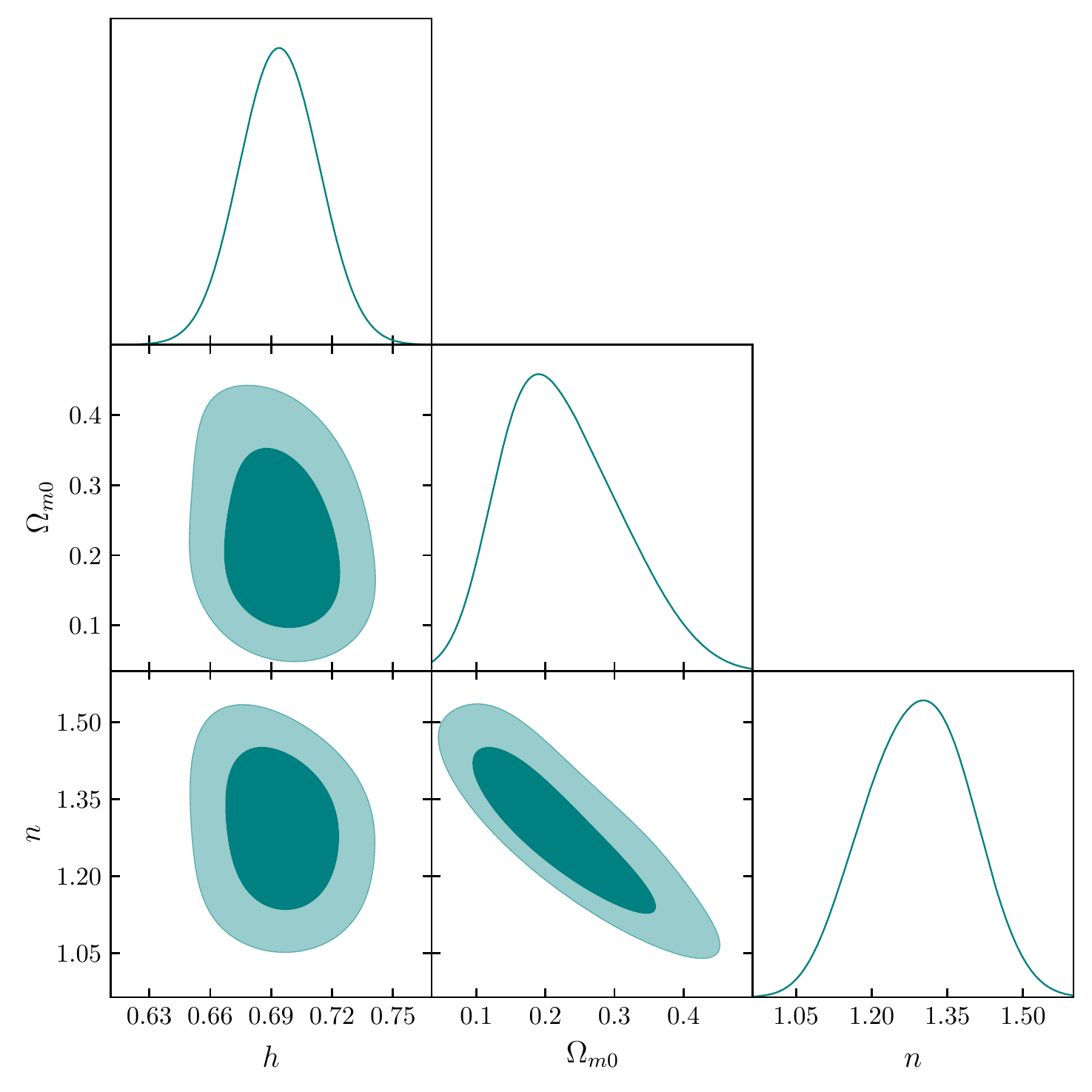}
    \caption{Marginalized 68\% and 95\% C.L. contours, with posterior distributions, for the free parameters of the $f(R,G)=R^nG^{1-n}$ model, as a result from the MCMC analysis of the combined SN+CC data.}
    \label{fig:contours}
\end{figure}

In Table~\ref{tab:results}, we report the $1\sigma$ and $2\sigma$ confidence level (C.L.) results of our MCMC analysis, while Fig.~\eqref{fig:contours} shows the 2-D marginalized contours and 1-D posterior distributions of the free parameters of the model.

\begin{table}
\begin{center}
\setlength{\tabcolsep}{1em}
\renewcommand{\arraystretch}{1.8}
\begin{tabular} {c c c c}
\hline
\hline
 Parameter & Mean & 68\% limits & 95\% limits \\
\hline
$h$  & 0.694 &$\pm \, 0.019 $ & $\pm \, 0.037$\\
$\Omega_{m0}$ & 0.223 & $^{+\,0.070}_{-\,0.098}$ & $^{+\,0.173}_{-\,0.152}$\\
$n$ & 1.29 & $_{-\,0.10}^{+\,0.11}$ & $^{+\,0.18}_{-\,0.19}$\\
\hline
\hline
\end{tabular}
\caption{Constraints at the 68\% and 95\% C.L. on the free parameters of the $f(R,G)=R^nG^{1-n}$ model, resulting from the MCMC analysis of the combined SN+CC data.}
\label{tab:results}
\end{center}
\end{table}

\subsection{Discussion of the results}
\begin{figure}
    \centering
    \includegraphics[width=3.3in]{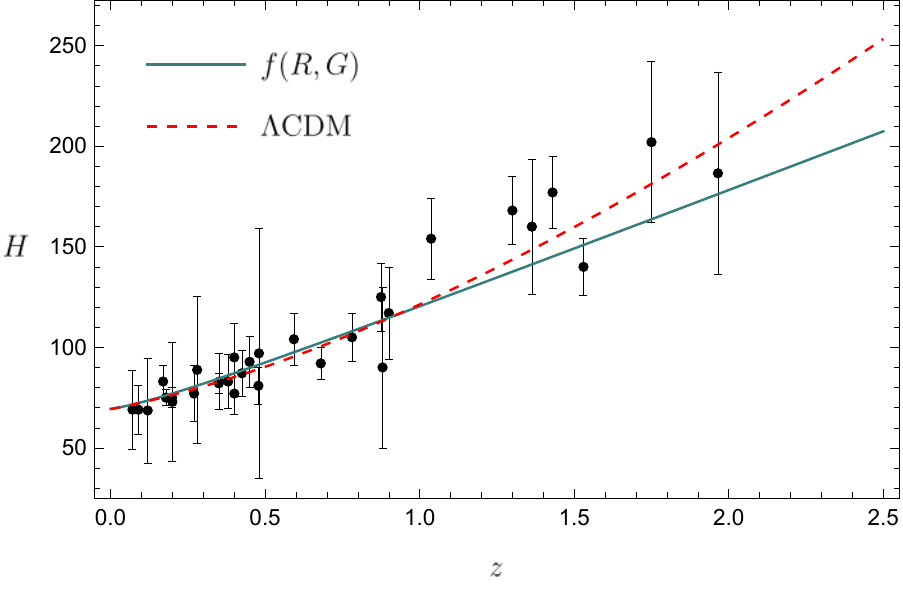}
    \caption{Behavior of the Hubble parameter as a function of the redshift for the $f(R,G)=R^nG^{1-n}$ model (teal) and the $\Lambda$CDM model (red). The cosmological parameters are fixed to the mean values obtained from our MCMC analysis (c.f. Table \ref{tab:results} and Eq.~\eqref{LCDM results}). The black dots and bars refer, respectively, to the best-fit and the $1\sigma$ uncertainties of the CC measurements considered in the present study.}
    \label{fig:LCDM comparison}
\end{figure}

\begin{figure}
    \centering
    \includegraphics[width=3.3in]{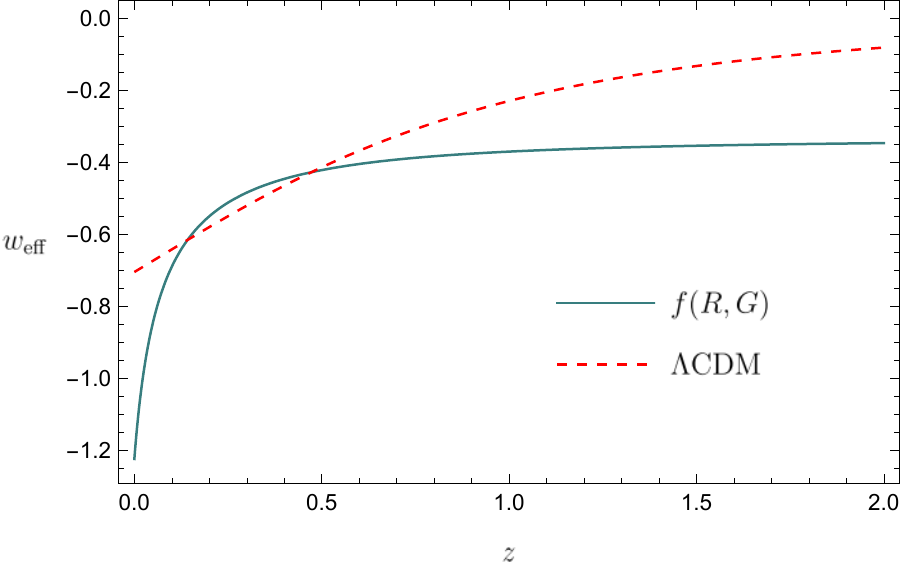}
    \caption{Comparison between the effective EoS parameter of the $f(R,G)=R^nG^{1-n}$ model (teal) and that of the $\Lambda$CDM model (red). The curves correspond to the mean results obtained from our MCMC analysis.}
    \label{fig:weff}
\end{figure}

Here, we shall discuss our findings by virtue of the predictions of the standard cosmological scenario. For this purpose, we recall the $1\sigma$ C.L. constraints on the $\Lambda$CDM model previously obtained from the MCMC analysis of the combined SN+CC data \cite{D'Agostino20}:
\begin{equation}
h=0.692 \pm 0.019\,, \quad \Omega_{m0}=0.296\,^{+\, 0.026}_{-\, 0.029}\,.
\label{LCDM results}
\end{equation}
From Table \ref{tab:results}, one can notice that the value of the Hubble constant resulting from the $f(R,G)$ model under consideration is fully consistent with the one predicted by $\Lambda$CDM. Our results differ by $\sim 1.7\sigma$ from the most recent (local) model-independent measurement by Riess et al. \cite{Riess22}, while agrees at $1\sigma$ with the estimate inferred by the Planck Collaboration \cite{Planck18}.

The constraints on the parameter $n$ (c.f. Table \ref{tab:results}) indicate more than $2\sigma$ deviations from the GR limit. As expected, the $f(R,G)$ model is capable of accounting for the dark energy effects without the need for the cosmological constant, due to the interplay between the Ricci scalar and the Gauss-Bonnet invariant.

Furthermore, although in agreement at the $1\sigma$ among each other, the mean result for the present matter density parameter is lower than both the late-time outcome given in (\ref{LCDM results}) and the early-time estimate of the Planck Collaboration assuming a $\Lambda$CDM cosmology, $\Omega_{m0}=0.315\pm 0.007$ \cite{Planck18}. The effect of such a discrepancy may be seen in Fig.~\ref{fig:LCDM comparison}, where we show the Hubble expansion rate of the $f(R,G)$ model compared to the $\Lambda$CDM prediction. Indeed, we note that the $f(R,G)$ model is able to reproduce fairly the accelerated behavior of the Universe up to $z\sim 1$. However, the differences between the two scenarios emerge as going backward in time, when the matter contribution starts becoming important until it eventually prevails over the dark energy effects.  Such behavior may translate into matter instabilities when density perturbations are taken into account during matter and radiation domination \cite{DeFelice10}.

The discrepancies with respect to the standard cosmological model are better visible from the analysis of the effective EoS parameter, given as
\begin{equation}
w_\text{eff}\equiv -1-\frac{2\dot{H}}{3H^2}=-1+\frac{2}{3}(1+z)\frac{H'}{H}\,.
\end{equation}
In Fig.~\ref{fig:weff}, in view of the mean results of our MCMC analysis, we can see that the effective EoS parameter of the $f(R,G)$ model shows a phantom behavior at the present time while, at high redshifts, it does not properly converge to zero as expected in order to have a standard matter-dominated phase. This is clearly due to the lower matter density abundance compared to $\Lambda$CDM, which strongly affects the cosmic evolution of $w_\text{eff}$.

\section{Energy conditions}
\label{sec:energy}
Starting from the expressions for the dark energy density and pressure given by Eqs.~\eqref{rhode} and \eqref{pide}, one can study the validity of the energy conditions associated with $f(R,G)= R^n G^{1-n}$ gravity.

The energy conditions play a fundamental role in defining physically viable models, especially in the context of extended theories of gravity (see \emph{e.g.}  \cite{Capozziello:2014bqa}).
They account for a set of inequalities the energy density and the pressure must satisfy, aiming to select the states of matter that are allowed in a given spacetime. Specifically, the null energy condition (NEC) imposes the trace of the energy-momentum tensor to be non-negative; the weak energy condition (WEC) is associated with the requirement of having positive energy; the dominant energy condition (DEC) validity implies that matter cannot travel faster than light, preserving the causality principle; finally, the strong energy condition (SEC) preserves the attractive nature of the gravitational field. 

Clearly, in GR, where the energy density and the pressure are those of standard matter, all the energy conditions are identically satisfied, whereas they can be violated as soon as exotic fluids are considered.
In the context of modified theories of gravity, the modified field equations can be recast such that the right-hand side can play the role of an effective energy-momentum tensor prompted by curvature. In this way, as shown in Eqs.~\eqref{rhode} and \eqref{pide}, the energy conditions can be also applied to the extra geometric terms that, in principle, can mimic the behavior of exotic matter fluids. Moreover, recasting $\rho_{de}$ and $p_{de}$ in terms of the cosmographic parameters, it is possible to determine the ranges of the free parameters of a given theory  leading to an accelerating cosmic expansion at late times. 

In our case, we study the behavior of the energy conditions depending on the free parameter $n$. As the values of the cosmographic parameters vary as a function of cosmic time, we shall consider their present-day estimates inferred from observations for finding theoretical bounds over $n$. These can be then confronted with the results of our analysis, to check for possible inconsistencies.
Specifically, using the values of the cosmographic parameters from the concordance $\Lambda$CDM model with $\Omega_{m0}=0.3$, namely $q_0=-0.55$, $j_0=1$ and $s_0=-0.35$ \cite{rocco_review}, it turns out that the energy conditions are satisfied in the following cases:
\begin{subequations}
\begin{align}
       \text{NEC}:& \  1 < n < 34\,,
\\
       \text{WEC}: & \  \nexists n \in \mathbb{R}\, ,
    \\
       \text{DEC}: & \  \nexists n \in \mathbb{R}\,,
    \\
      \text{SEC}: & \  1 < n < 34\,.
\end{align}
\end{subequations}
We notice that the WEC and DEC are identically violated for any values of $n$, meaning that the extra geometric terms may give rise to negative effective pressure and to the violation of the causality principle. On the other hand, our constraints on $n$ are in agreement with the range admitted for the validity of the WEC and DEC.

A further possibility may be to consider purely model-independent estimates of the cosmographic parameters. In particular, using the recent findings of \cite{Capozziello20}, namely $q_0=-0.6$, $j_0=1.32$ and $s_0=8.47$, we obtain
\begin{subequations}
\begin{align}
       \text{NEC}:& \ -29 < n < 1\,,
    \\
       \text{WEC}:& \ -29 < n < 1\,,
    \\
       \text{DEC}:& \ -29 < n < -0.25\,,
    \\
       \text{SEC}:& \ -19 < n < 1\,.
\end{align}
\end{subequations}
This case is of particular interest since these values have been obtained through a kinematic procedure that does not rely on any \emph{a priori} assumed cosmological model.
It is worth noting that, in all cases, our constraints over $n$ violate all the energy conditions, thus mimicking a dark fluid behavior.

\section{Final remarks and perspectives}
\label{conclSec}

In this work, we studied the cosmological behavior of a specific class of modified Gauss-Bonnet gravity models. To this aim, we first outlined the main properties of a gravitational action involving a general combination of the Ricci scalar and the Gauss-Bonnet invariant.
Hence, assuming a flat FLRW cosmological background, we obtained the point-like Lagrangian of the theory and the related equations of motion. 
We then applied the Noether symmetry approach to select viable functions and reduce the dimension of the minisuperspace, thus allowing us to find exact solutions to the vacuum field equations. The selected function, namely $f(R,G) = R^n G^{1-n}$, reduces to GR as soon as the real constant $n$ approaches the unity. However, the scenario under study does not recover the cosmological constant case explicitly, so it is particularly interesting toward finding viable alternatives to the standard $\Lambda$CDM model, capable of mimicking the dark energy behavior and avoiding the conceptual issues proper of $\Lambda$.
In our case, we showed that the right-hand sides of the modified Friedmann equations can be understood as effective energy density and pressure due to curvature. We thus found the expression of the dark energy EoS parameter in terms of both the cosmographic parameters and the free constant of the theory, $n$.

Furthermore, we investigated the cosmological features of the $f(R,G)$ model in the presence of matter fields. Assuming non-relativistic pressureless matter and neglecting the late-time contribution of the radiation fluid, we numerically solved the first Friedmann equation to find the redshift behavior of the Hubble parameter. In so doing, we considered the $\Lambda$CDM model to find suitable initial conditions over $H(z)$ and its derivatives. Then, we employed the most recent low-redshift observations to directly compare our theory with the model-independent predictions of the cosmic expansion. In particular, we performed a Bayesian analysis through the MCMC method, using the combination of Supernovae Ia and Hubble observational data. Assuming uniform prior distributions, we obtained constraints over the free parameters of the model at the $1\sigma$ and $2\sigma$ C.L., which allowed us to reconstruct the cosmological evolution of the Hubble expansion rate and the total effective EoS parameter. Our analysis shows that the $f(R,G)$ model is able to explain the current acceleration of the Universe without resorting to $\Lambda$. However, a close comparison with the predictions of the standard cosmological scenario reveals that the $f(R,G)$ model starts to considerably deviate from $\Lambda$CDM as the redshift increases, thus failing to provide a standard matter-dominated era. This result is confirmed by the behavior of the effective EoS parameter, which does not vanish when $z\gg 1$. This appears to be common with other modified gravity theories, such as $f(R)$, where matter instabilities occur as density perturbations are taken into account.

Finally, we complemented our analysis by studying the validity of the energy conditions, when written in terms of effective pressure and energy density. Specifically, we considered two different sets of cosmographic parameters, namely the values inferred from the concordance $\Lambda$CDM model and those emerging from a kinematic model-independent approach to the dark energy problem. In the first case, we showed that the WEC and DEC are identically violated for any $n$, while the NEC and SEC are satisfied for $1<n<34$. Therefore, the value $n \sim 1.29$ obtained from our observational analysis lies within the validity ranges of NEC and SEC. On the other hand, considering the second set of cosmographic parameters, it turns out that the NEC, WEC and SEC are satisfied for $n<1$, whereas the DEC is fulfilled for $n<0.25$. It is worth stressing that, in the latter case, the value of $n$ selected by the cosmological analysis violates all the energy conditions, confirming that the $f(R,G)$ model is capable of behaving like GR with the cosmological constant, thus mimicking the dark energy features.

To conclude, the model under investigation well behaves when confronted with observations at late times, though it is unable to properly address the matter-dominated epoch. Nonetheless, similarly to other modified gravity models, a typical solution  to the latter problem consists of considering the action of screening mechanisms, implying 
a gravitational Lagrangian characterized by the presence of additional coupling constants, whose contributions become dominant at different spatial/temporal scales. This, in principle, could allow to recover the standard behavior at intermediate redshifts and thus properly predict the formation of cosmic structures. In this respect, useful insights could arise from the study of cosmological perturbations and  the comparison with the growth of matter overdensity measurements.

\begin{acknowledgements}

The authors acknowledge the support of Istituto Nazionale di Fisica Nucleare (INFN), {\it iniziative specifiche} GINGER and QGSKY. The authors would also like to thank Salvatore Capozziello for useful discussions.

\end{acknowledgements}

\appendix
\section{Useful relations}
\label{app:relations}

For the sake of completeness, we here report some useful relations for determining the cosmological dynamics in the case of $f(R,G) = R^n G^m$ gravity. Specifically, starting from the definitions given in Eqs.~\eqref{cosmoexprR} and \eqref{cosmoexprG}, the time derivatives of the Ricci scalar and the Gauss-Bonnet term take the form
\begin{align}
\dot{R}&= 6(4H\dot{H}+\ddot{H})\,,\\
\dot{G}&=24H(4H^2\dot{H}+2\dot{H}^2+H\ddot{H})   \,.
\end{align}

Moreover, the time derivatives of the functions appearing in Eqs.~\eqref{rde} and \eqref{pde} can be expressed in terms of the above equations and the derivatives with respect to $R$ and $G$ as follows:
\begin{align}
\dot{f_R}=&\ \dot R f_{RR}+\dot G f_{RG}\,,\\
 \dot{f_G}=&\ \dot G f_{GG} + \dot R f_{RG}\,,\\
 \ddot{f_R}=&\ \ddot{R}f_{RR}+\ddot{G}f_{RG}+\dot{R}^2 f_{RRR}+\dot{G}^2 f_{RGG}+2\dot{R}\dot{G}f_{RRG}\,,\\
\ddot{f_G}=&\ \ddot{G} f_{GG} +\ddot R f_{RG}+\dot{G}^2 f_{GGG}+\dot{R}^2 f_{RRG}+2\dot{R}\dot{G}f_{RGG}\,.
\end{align}

\section{Effective dark energy pressure}
\label{app:p_de}

In the case of $f(R,G) = R^n G^m$ gravity models,
the dark energy pressure \eqref{pde} can be written in the compact form \eqref{eq:p_de}, where the explicit expressions of the coefficients $c_k(j,s;n,m)$ are
\begin{align}
c_0=&\ m  \left(-m ^2+3 m -2\right)j^2\,,
\\
c_1=&\ m (m -1)  \left[3 j^2 (n +m -2)-j (4 n +6 m -6)+s\right],
\end{align}
\begin{align}
c_2=&\ m \left\{2 j \left[9 n  m +n  (4 n -11)+7 m ^2-16 m +9\right]\right. \nonumber \\
&\hspace{0.6cm}-(2 n +3 m -3) (2 n +3 m +s-2)\nonumber  \\
&\left.\hspace{0.6cm}-3 j^2 (n +m-2) (n +m -1)\right\},
\\
c_3=&\  (m -1) \left[m ^2 \left(j^2-6 j+15\right)-m  \left(2 j^2-18 j-3 s+15\right)\right.\nonumber \\
&\left.+3\right]+n ^2 \left[12 m +3 (m -1) j^2-2 (5 m -7) j+s-10\right]\nonumber \\ 
& +(j-2)^2 n^3+n  \left[19 m ^2-25 m +\left(3 m ^2-6 m +2\right) j^2\right.\nonumber \\
&\left.-2 \left(8 m ^2-13 m +5\right) j+4 m  s-s+3\right], 
\\
c_4=&\ n \left\{3 m -2 \left[m  (3 m +2) j+j+m  (s-3 m )\right]+s+13\right\}\nonumber \\
&+2 n ^3 (2-j)-n ^2 \left[-5 m +2 (m -2) j+s+8\right] \nonumber \\
&+(m -1) \left\{9+m  \left[5 m -6 (m +1) j-s+11\right]\right\},\\
c_5=&\ n ^2 \left[m  (4 j-9)-8\right]+(m -1) \left[m ^2 (4 j-15)-3 m-9\right] \nonumber \\
&+ n  \left(m ^2 (8 j-17)+m  (6-4 j)-2\right)+n^3\,, 
\\
c_6=&\ 4n m (2-n)  -(n -4) n +3 m ^2-3\,,
\\
c_7=&\ 2 m  (2 m -1) (n +m -1) \,.
\end{align}

\section{Solutions to vacuum field equations}
\label{app:vacuum solutions}

Making use of the relations reported in Appendix~\ref{app:relations}, it is possible to find analytic solutions for the scale factor of $f(R,G)=R^n G^m$ gravity in vacuum.
In particular, it turns out that the theory under consideration admits two different sets of solutions. The first one is a time power-law scale factor of the form $a(t) = a_0 t^\ell$, with 
\begin{eqnarray}
&&\small \ell =  \left[\frac{1}{{2 (2 m+n-2)}}\right]\Big\{-8 m^2-8 m n+11 m-2 n^2  \nonumber
\\
&&+4 n-3 \pm \Big[\Big(8 m^2+8 m n-11 m+2 n^2  -4 n+3\Big)^2 
   \\
&& +4 (2 m+n-2) \Big(4 m^2+6 m
   n-5 m +2 n^2-3 n+1 \Big) \Big]^{\frac{1}{2}}   \Big\}. \nonumber
\end{eqnarray}
Setting $m = 1-n$, the solution takes the form $a(t) = a_0 t^{2n-1}$, as written in Eq.~\eqref{ExactSol}. Another solution occurs when considering exponential scale factors of the form $a(t) = a_0 e^{s\, t}$, with $s$ being a real number. However, in order for this scale factor to be the solution to the field equation, we must also have $m=1-n/2$.

\section*{Declarations}
\textbf{Fundings}. The authors received no financial support for the research, authorship, and/or publication of this article.
\\ \\
\textbf{Data Availability Statement}. The manuscript does not contain any material from third parties; all of the material is owned by the authors and/or no permissions are required.
\\ \\
\textbf{Author Contributions Statement}. The authors equally contributed to the conceptualization, analysis and writing of the manuscript.
\\ \\
\textbf{Conflict of Interest}. The authors declare that they have no competing interests as defined by Springer, or other interests that might be perceived to influence the results and/or discussion reported in this paper.


\begin{thebibliography}{99}

\bibitem{Riess98}
A. G. Riess et al., Astron. J. \textbf{116}, 1009 (1998).

\bibitem{Perlmutter99}
S. Perlmutter et al., Astrophys. J. \textbf{517},  565  (1999).

\bibitem{Peebles03}
P. J. E. Peebles and B. Ratra, Rev. Mod. Phys. \textbf{75},  559 (2003).

\bibitem{Copeland06}
E. Copeland, M. Sani and S. Tsujikawa,  Int. J. Mod. Phys. D \textbf{15}, 1753 (2006).

\bibitem{Weinberg89}
S. Weinberg, Rev. Mod. Phys. \textbf{61}, 1 (1989).

\bibitem{Padmanabhan03}
T. Padmanabhan,  Phys. Rept. \textbf{380},  235 (2003).

\bibitem{Carroll01}
S. M. Carroll,  Living Rev. Rel. \textbf{4}, 1 (2001).

\bibitem{D'Agostino22}
R.~D'Agostino, O.~Luongo and M.~Muccino,
Class. Quant. Grav. \textbf{39}, 195014 (2022).



\bibitem{Ratra88}
B. Ratra and P. J. E. Peebles, Phys. Rev. D \textbf{37}, 3406 (1988).

\bibitem{Caldwell98}
R. R. Caldwell, R. Dave and P. J. Steinhardt, Phys. Rev. Lett. \textbf{80}, 1582 (1998).

\bibitem{Zlatev99}
I. Zlatev, L. Wang and P. J. Steinhardt, Phys. Rev. Lett. \textbf{82}, 896 (1999). 

\bibitem{Sahni00}
V. Sahni and L. M. Wang,  Phys. Rev. D \textbf{62}, 103517 (2000).

\bibitem{Scherrer04}
R. J. Scherrer, Phys. Rev. Lett. \textbf{93}, 011301 (2004).

\bibitem{Capozziello06}
S. Capozziello, S. Nojiri and S. D. Odintsov, Phys. Lett. B \textbf{632}, 597 (2006).


\bibitem{Anton-Schmidt}
S. Capozziello, R. D'Agostino and O. Luongo, Phys. Dark Univ. \textbf{20}, 1 (2018);  
S. Capozziello, R. D'Agostino, R. Giamb\`o and O. Luongo, Phys. Rev. D \textbf{99}, 023532 (2019); 
K. Boshkayev, R. D'Agostino and O. Luongo, Eur. Phys. J. C \textbf{79}, 332 (2019).

\bibitem{D'Agostino22b}
R. D'Agostino and O. Luongo, Phys. Lett. B \textbf{829}, 137070 (2022).


\bibitem{Clifton:2011jh}
T.~Clifton, P.~G.~Ferreira, A.~Padilla and C.~Skordis,
Phys. Rept. \textbf{513}, 1-189 (2012)

\bibitem{Nojiri:2017ncd}
S.~Nojiri, S.~D.~Odintsov and V.~K.~Oikonomou,
Phys. Rept. \textbf{692}, 1 (2017).


\bibitem{Carroll04}
S. M. Carroll, V. Duvvuri, M. Trodden and M. S. Turner, Phys. Rev. D \textbf{70}, 043528 (2004).

\bibitem{Starobinsky07}
A. A. Starobisnky, JETP Lett. \textbf{86}, 157 (2007).

\bibitem{Nojiri11}
S. Nojiri and S. Odintsov,  Phys. Rept. \textbf{505}, 59 (2011). 

\bibitem{Joyce:2014kja}
A.~Joyce, B.~Jain, J.~Khoury and M.~Trodden,
Phys. Rept. \textbf{568}, 1-98 (2015)

\bibitem{Koyama:2015vza}
K.~Koyama,
Rept. Prog. Phys. \textbf{79}, no.4, 046902 (2016)

\bibitem{Amendola:2006kh}
L.~Amendola, D.~Polarski and S.~Tsujikawa,
Phys. Rev. Lett. \textbf{98}, 131302 (2007)

\bibitem{Sotiriou10}
T. P. Sotiriou and V. Faraoni, Rev. Mod. Phys. \textbf{82}, 451 (2010).

\bibitem{Dolgov:2003px}
A.~D.~Dolgov and M.~Kawasaki,
Phys. Lett. B \textbf{573}, 1-4 (2003)

\bibitem{Faraoni:2006sy}
V.~Faraoni,
Phys. Rev. D \textbf{74}, 104017 (2006)

\bibitem{DeFelice_review}
A. De Felice and  S. Tsujikawa,  Living Rev. Rel. \textbf{13}, 3 (2010).

\bibitem{Capozziello_review}
S. Capozziello and M. De Laurentis,  Phys. Rept. \textbf{509},  167 (2011).


\bibitem{Nojiri:2005jg}
S.~Nojiri and S.~D.~Odintsov,
Phys. Lett. B \textbf{631}, 1 (2005).

\bibitem{Li:2007jm}
B.~Li, J.~D.~Barrow and D.~F.~Mota,
Phys. Rev. D \textbf{76}, 044027 (2007).


\bibitem{Elizalde:2010jx}
E.~Elizalde, R.~Myrzakulov, V.~V.~Obukhov and D.~Saez-Gomez,
Class. Quant. Grav. \textbf{27}, 095007 (2010).

\bibitem{DeFelice:2009aj}
A.~De Felice and S.~Tsujikawa,
Phys. Rev. D \textbf{80}, 063516 (2009).

\bibitem{Oikonomou:2022ksx}
V.~K.~Oikonomou, P.~D.~Katzanis and I.~C.~Papadimitriou,
Class. Quant. Grav. \textbf{39} (2022) no.9, 095008

\bibitem{Oikonomou:2021kql}
V.~K.~Oikonomou,
Class. Quant. Grav. \textbf{38} (2021) no.19, 195025

\bibitem{Odintsov:2020vjb}
S.~D.~Odintsov, V.~K.~Oikonomou, F.~P.~Fronimos and K.~V.~Fasoulakos,
Phys. Rev. D \textbf{102} (2020) no.10, 104042

\bibitem{Bajardi:2021hya}
F.~Bajardi, D.~Vernieri and S.~Capozziello,
J. Cosm. Astrop. Phys. \textbf{11}, 057 (2021).

\bibitem{Lovelock:1971yv}
D.~Lovelock,
J. Math. Phys. \textbf{12}, 498 (1971).

\bibitem{Mardones:1990qc}
A.~Mardones and J.~Zanelli,
Class. Quant. Grav. \textbf{8}, 1545 (1991).

\bibitem{Achucarro:1986uwr}
A.~Achucarro and P.~K.~Townsend,
Phys. Lett. B \textbf{180}, 89 (1986).

\bibitem{Gomez:2011zzd}
F.~Gomez, P.~Minning and P.~Salgado,
Phys. Rev. D \textbf{84}, 063506 (2011).

\bibitem{Leigh:1989jq}
R.~G.~Leigh,
Mod. Phys. Lett. A \textbf{4}, 2767 (1989).

\bibitem{Tseytlin:1997csa}
A.~A.~Tseytlin,
Nucl. Phys. B \textbf{501}, 41 (1997).

\bibitem{Marathe:1989tm}
K.~B.~Marathe and G.~Martucci,
J. Geom. Phys. \textbf{6}, 1-106 (1989)

\bibitem{Bajardi:2020osh}
F.~Bajardi and S.~Capozziello,
Eur. Phys. J. C \textbf{80}, 704 (2020).


\bibitem{DeFelice}
A.~De Felice, J.~M.~Gerard and T.~Suyama,
Phys. Rev. D \textbf{82}, 063526 (2010);
A.~De Felice, T.~Suyama and T.~Tanaka,
Phys. Rev. D \textbf{83}, 104035 (2011).

\bibitem{Sadjadi:2010kp}
H.~M.~Sadjadi,
EPL \textbf{92}, 50014 (2010).

\bibitem{Makarenko:2012gm}
A.~N.~Makarenko, V.~V.~Obukhov and I.~V.~Kirnos,
Astrophys. Space Sci. \textbf{343}, 481 (2013).

\bibitem{DeLaurentis:2013ska}
M.~De Laurentis and A.~J.~Lopez-Revelles,
Int. J. Geom. Meth. Mod. Phys. \textbf{11}, 1450082 (2014).

\bibitem{Elizalde:2020zcb}
E.~Elizalde, S.~D.~Odintsov, V.~K.~Oikonomou and T.~Paul,
Nucl. Phys. B \textbf{954}, 114984 (2020).


\bibitem{Mustafa:2020jln}
G.~Mustafa, M.~F.~Shamir and X.~Tie-Cheng,
Phys. Rev. D \textbf{101}, 104013 (2020).

\bibitem{deMartino:2020yhq}
I.~de Martino, M.~De Laurentis and S.~Capozziello,
Phys. Rev. D \textbf{102}, 063508 (2020).


\bibitem{Nojiri:2022xdo}
S.~Nojiri, S.~D.~Odintsov and T.~Paul,
Phys. Dark Univ. \textbf{35}, 100984 (2022).

\bibitem{Akbarieh:2021vhv}
A.~R.~Akbarieh, S.~Kazempour and L.~Shao,
Phys. Rev. D \textbf{103}, 123518 (2021)

\bibitem{Carroll:2004de}
S. M. Carroll, A. De Felice, V. Duvvuri, D. A. Easson, M. Trodden and M. S. Turner, Phys. Rev. D \textbf{71}, 063513 (2005).

\bibitem{Bogdanos:2009tn}
C.~Bogdanos, S.~Capozziello, M.~De Laurentis and S.~Nesseris,
Astropart. Phys. \textbf{34}, 236 (2010).

\bibitem{DeFelice:2009ak}
A.~De Felice and T.~Suyama,
J. Cosm. Astrop. Phys. \textbf{06}, 034 (2009).

\bibitem{Astashenok:2015haa}
A.~V.~Astashenok, S.~D.~Odintsov and V.~K.~Oikonomou,
Class. Quant. Grav. \textbf{32}, 185007 (2015).

\bibitem{Nojiri:2018ouv}
S.~Nojiri, S.~D.~Odintsov and V.~K.~Oikonomou,
Phys. Rev. D \textbf{99}, 044050 (2019).

\bibitem{Acunzo:2021gqc}
A.~Acunzo, F.~Bajardi and S.~Capozziello,
Phys. Lett. B \textbf{826}, 136907 (2022).

\bibitem{Bajardi:2021tul}
F.~Bajardi and S.~Capozziello,
Int. J. Geom. Meth. Mod. Phys. \textbf{18}, 2140002 (2021).

\bibitem{Capozziello:2014ioa}
S.~Capozziello, M.~De Laurentis and S.~D.~Odintsov,
Mod. Phys. Lett. A \textbf{29}, 1450164 (2014).

\bibitem{Camci:2018apx}
U.~Camci,
Symmetry \textbf{10}, 719 (2018).

\bibitem{Bajardi:2020xfj}
F.~Bajardi and S.~Capozziello,
Int. J. Mod. Phys. D \textbf{29}, 2030015 (2020).

\bibitem{Urban:2020lfk}
Z.~Urban, F.~Bajardi and S.~Capozziello,
Int. J. Geom. Meth. Mod. Phys. \textbf{17}, 2050215 (2020).

\bibitem{Dialektopoulos:2018qoe}
K.~F.~Dialektopoulos and S.~Capozziello,
Int. J. Geom. Meth. Mod. Phys. \textbf{15}, 1840007 (2018).

\bibitem{Bajardi:2022ypn}
F.~Bajardi and S.~Capozziello,
Cambridge University Press, 2022,
doi:10.1017/9781009208727

\bibitem{Weinberg72}
S. Weinberg, \emph{Gravitation and cosmology: Principles and applications of the general theory of relativity}, Wiley, New York (1972).

\bibitem{Visser05}
M. Visser, Gen. Rel. Grav. \textbf{37}, 1541 (2005).

\bibitem{rocco_chebyshev}
S. Capozziello, R. D'Agostino and O. Luongo, Mon. Not. Roy. Astron. Soc. \textbf{476}, 3924 (2018).

\bibitem{SantosDaCosta:2018bbw}
S.~Santos Da Costa, F.~V.~Roig, J.~S.~Alcaniz, S.~Capozziello, M.~De Laurentis and M.~Benetti,
Class. Quant. Grav. \textbf{35}, 075013 (2018).
 
\bibitem{Bamba:2010wfw}
K.~Bamba, S.~D.~Odintsov, L.~Sebastiani and S.~Zerbini,
Eur. Phys. J. C \textbf{67}, 295 (2010).

\bibitem{DeLaurentis:2015fea}
M.~De Laurentis, M.~Paolella and S.~Capozziello,
Phys. Rev. D \textbf{91}, 083531 (2015).

\bibitem{Bahamonde:2019swy}
S.~Bahamonde, K.~Dialektopoulos and U.~Camci,
Symmetry \textbf{12}, 68 (2020).

\bibitem{Scolnic18}
D. M. Scolnic et al., Astrophys. J. \textbf{859}, 101 (2018).


\bibitem{Capozziello18}
S. Capozziello, R. D'Agostino and O. Luongo, Phys. Dark Univ. \textbf{20}, 1 (2018).

\bibitem{D'Agostino18}
R. D'Agostino and O. Luongo, Phys. Rev. D \textbf{98}, 124013 (2018).

\bibitem{D'Agostino19}
R. D'Agostino, Phys. Rev. D \textbf{99}, 103524 (2019).


\bibitem{Guy07}
J. Guy et al., Astron. Astrophys. \textbf{466}, 11 (2007).

\bibitem{Betoule14}
M. Betoule et al., Astron. Astrophys. \textbf{568}, A22 (2014).

\bibitem{Riess18}
A. G. Riess et al., Astrophys. J. \textbf{853}, 126 (2018).


\bibitem{DAgostino:2022tdk}
R.~D'Agostino and R.~C.~Nunes,
Phys. Rev. D \textbf{106}, 124053 (2022).

\bibitem{Jimenez02}
R. Jimenez and A. Loeb, Astrophys. J. \textbf{573}, 37 (2002).

\bibitem{Hastings70}
W. K. Hastings, Biometrika \textbf{57}, 97 (1970). 

\bibitem{D'Agostino20}
R. D'Agostino and R. C. Nunes, Phys. Rev. D \textbf{101}, 103505 (2020).

\bibitem{Riess22}
A. G. Riess et al., Astrophys. J. Lett. \textbf{934}, L7 (2022).

\bibitem{Planck18}
Planck Collaboration (N. Aghanim et al.),  Astron. Astrophys. \textbf{641}, A6 (2020), Astron. Astrophys. \textbf{652}, C4 (erratum) (2021). 

\bibitem{DeFelice10}
A. De Felice, D. F. Mota and S. Tsujikawa, Phys. Rev. D \textbf{81}, 023532 (2010).

\bibitem{Capozziello:2014bqa}
S.~Capozziello, F.~S.~N.~Lobo and J.~P.~Mimoso,
Phys. Rev. D \textbf{91}, 124019 (2015).

\bibitem{rocco_review}
S. Capozziello, R. D'Agostino and O. Luongo, Int. J. Mod. Phys. D \textbf{28}, 1930016 (2019).

\bibitem{Capozziello20}

S. Capozziello, R. D'Agostino and O. Luongo, Mon. Not. Roy. Astron. Soc. \textbf{494}, 2576 (2020).

\end{thebibliography}
\end{document}